\documentclass[aps,prl,twocolumn,groupedaddress]{revtex4}
\usepackage{graphicx}

\newcommand{\xdir}{$[ 1\overline{1} 0 ]$}
\newcommand{\ydir}{$[ 110 ]$}

\newcommand{\Vdc}{$V_{dc}$}
\newcommand{\Vac}{$V_{ac}$}
\newcommand{\Idc}{$I_{dc}$}

\newcommand{\Vavg}{$\langle V_{rms} \rangle$}

\newcommand{\IV}{$I$-$V$}

\begin{document}

\title{Observation of narrow-band noise accompanying the breakdown of 
insulating states in high Landau levels}

\author{K.~B.~Cooper$^1$, J.~P. Eisenstein$^1$, 
L.~N. Pfeiffer$^2$, and K. W. West$^2$}

\affiliation{$^1$California Institute of Technology, Pasadena CA 91125 
\\
	 $^2$Bell Laboratories, Lucent Technologies, Murray Hill, NJ 
07974\\}


\begin{abstract} 
Recent magnetotransport experiments on high mobility two-dimensional
electron systems have revealed many-body electron states unique to high Landau levels. Among these are re-entrant integer quantum Hall states which undergo sharp transitions to conduction above some threshold field. Here we report that these transitions are often accompanied by narrow- and broad-band noise with frequencies which are strongly dependent on the magnitude of the applied dc current. 

\end{abstract}


\maketitle

Strong evidence now exists which suggests that in the presence of a modest perpendicular 
magnetic field $B$, a clean two-dimensional electron system (2DES) with density $n_s$ supports many-body phases distinct from the 
fractional quantum Hall effect (FQHE) states which dominate at high fields\cite{jpesscom}.  These new phases form when the 2DES occupies three or 
more orbital Landau levels (LLs).  Near half filling of the spin resolved LLs, e.g. at filling fraction $\nu=hn_s/eB=9/2$, 11/2, etc., states exhibiting highly anisotropic longitudinal resistance develop at temperatures below about 100 mK\cite{lilly1,du}.  These are believed to be closely related to the 
striped charge density wave (CDW) states predicted by Hartree-Fock theory\cite{KFS,MC}. 

Near 1/4 and 3/4 filling of the same high LLs, additional phases exist whose main experimental signature is an {\it isotropic} vanishing of the longitudinal 
resistance and a re-entrant integer quantization of the Hall resistance\cite{lilly1,du}. 
These states are separated from the nearby conventional integer quantum Hall states by narrow regions of non-zero longitudinal resistance and non-quantized Hall resistance.  The re-entrant aspect of these states suggests that 
they reflect collective, as opposed to single particle, insulating 
configurations of the electrons in the uppermost LL. Indeed, prevailing theory 
predicts that the ground state of the 2DES at these fillings is a pinned triangular CDW, analogous to a Wigner crystal but with two or more electrons per unit cell\cite{KFS,MC,rezayi2,yoshi}. Estimates of the lattice constant of such 
``bubble" phases are on the order of 100 nm under typical circumstances. This 
picture is supported by non-linear dc current-voltage (\IV\/) 
measurements in which sharp onsets to conduction are observed in the re-entrant integer quantum Hall effect (RIQHE) states\cite{cooper1}. Such threshold behavior is a common feature of transport in conventional CDW systems, such as NbSe$_3$\cite{gruner}. It has also been observed in 2DESs at low filling of the $N=0$ LL where Wigner crystallization is thought to occur\cite{wigner1}.  Recently, Lewis \emph{et al.}\cite{lewis} have observed sharp resonances in the microwave conductivity of the 2DES in the vicinity of the RIQHE states and have suggested that they may reflect a pinning mode of the bubble crystal\cite{resonances}.

Here we report a remarkable new aspect of the RIQHE: the abrupt onset of conduction at high dc bias is often accompanied by narrow- and broad-band 
noise which is very sensitive to the dc current carried by the 2DES. While it is 
tempting to compare the narrow-band noise to the ``washboard" noise 
encountered in the sliding transport of conventional CDW compounds\cite{gruner}, the low frequencies ($\sim 5~{\rm kHz}$) we observe may point to a different origin.  Nevertheless, the dynamical features described here form an important part of the phenomenology of the RIQHE and include the first observation of narrow-band noise in the quantum Hall regime\cite{bbnoise}.

Noise accompanying RIQHE breakdown has been observed in four different GaAs/AlGaAs heterostructures. We focus here on a 30 nm GaAs quantum well containing a 2DES with density $n_s = \rm 2.9\times10^{11}~cm^{-2}$ and mobility $\mu =\rm 2.3\times10^7~cm^2/Vs$. Eight diffused In ohmic contacts were positioned at the corners and side midpoints of the $4 \times 4$ mm square sample. The linear response longitudinal resistance was obtained by driving low level purely ac current (typically 10 nA at 13 Hz) between midpoint contacts on opposite sides of the sample and recording the voltage drop between adjacent corner contacts. The non-linear response of the sample was observed using purely dc current excitation and monitoring both the dc and ac components of the resulting voltage. 

Figure~1a shows the linear response longitudinal resistances measured with net 
current flow along the orthogonal \ydir\ and \xdir\ crystal axes at $T=50$ mK over the filling factor range $5>\nu>4$. As is usually the case when no external 
symmetry-breaking field is present\cite{note1}, the resistance shows a deep 
minimum around $\nu=9/2$ for current flow along \ydir\ and a strong maximum for 
current flowing along \xdir. Flanking the region of anisotropy are the RIQHE 
states around $\nu \approx 4+1/4$ and 4+3/4 filling, where the longitudinal resistance vanishes isotropically.  Although not shown, the Hall resistance is quantized in these regions, at $h/4e^2$ and $h/5e^2$ respectively. 

\IV\ characteristics of the RIQHE states were obtained using a circuit
shown in Fig.~1b. At $\nu=4.27$, near the center
of the high-field RIQHE, \Vdc\ versus \Idc\ is shown in Fig.~1b. For
small \Idc\ the RIQHE remains intact, i.e. $V_{dc}=0$. The current is
flowing entirely within the filled $N=0$ and $N=1$ lower LLs while the
quasiparticles in the partially-filled $N=2$ LL remain insulating. At
$I_{dc} \approx 1.1~\mu \rm{A}$, however, \Vdc\ abruptly becomes
non-zero as the RIQHE breaks down and current begins to flow in the
$N=2$ LL\cite{cooper1}. 

Fig.~1c shows that along with the sharp dc response, a nonzero \Vac\ appears as the RIQHE breaks down. Four 10 ms traces of the net longitudinal voltage \Vac+\Vdc, taken at different values of \Idc, are displayed. For \Idc$=0.5$ $\mu$A, \Vac\ is very small and displays only background noise.
Above the dc threshold current of 1.1 $\mu$A, however, \Vac\ begins to fluctuate. Immediately above threshold, at \Idc$=1.2$ $\mu$A, a reproducible 
burst-like pattern separated by irregular intervals is observed. At higher currents \Vac\ shows unambiguous narrow-band noise, oscillating around 3 kHz for \Idc$=1.4$ $\mu$A. As \Idc\ is increased further, the noise frequency rises: by \Idc$=1.7$ $\mu$A, fluctuations on time scales of less than 0.1 ms are seen. We emphasize that this noise in \Vac\ develops spontaneously in the RIQHE regions of high LLs under purely dc current excitation. The noise arises from the 2DES itself and is not due to interactions with the external circuitry.

\begin{figure}
\centering
\includegraphics[width=3.25 in,bb=107 103 511 687]{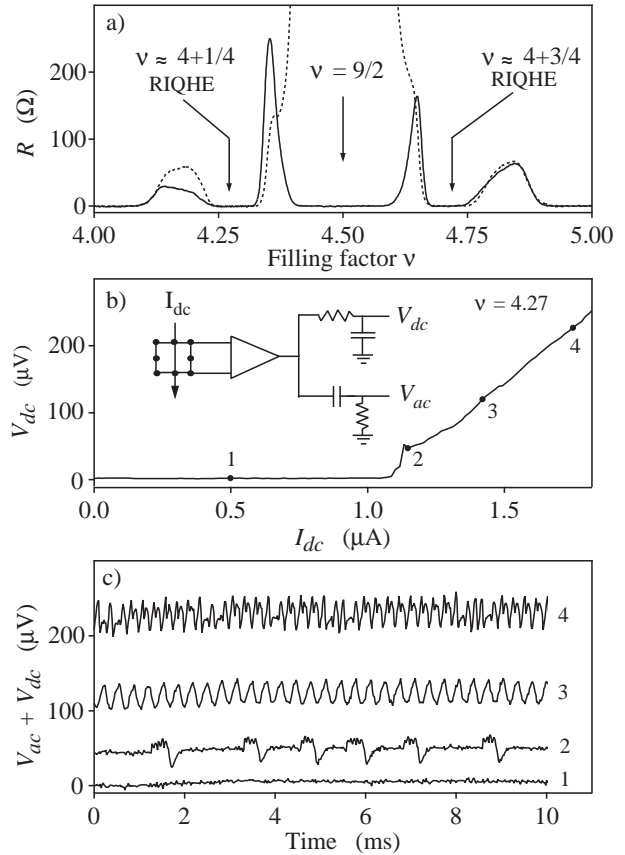}
\caption{\label{fig:fig1} Overview of the RIQHE and its breakdown 
phenomena at $T=50$ mK.
a) Linear response longitudinal resistance with the current directed along \ydir\ (solid) and \xdir\ (dashed).  The RIQHE states straddle the region of anisotropic transport around $\nu=9/2$. b) Non-linear \IV\ data taken near the center of the $\nu \approx 4+1/4$ RIQHE. c) Time dependences of the net longitudinal voltage \Vac+\Vdc, taken at the values of \Idc\ labeled in b). 
}
\end{figure}

Figure~2 illustrates the spectral content of the noise associated with the breakdown of the RIQHE near $\nu \approx 4+1/4$.  Figure 2a shows a typical dc \IV\ characteristic, the threshold to conduction occuring close to $1.0~\mu \rm{A}$, while Fig. 2b displays a logarithmic plot of the simultaneously recorded Fourier power spectrum of \Vac. The full gray scale in Fig. 2b corresponds to a noise power variation of 37 dB. The faint horizontal lines running across the entire image are artifacts arising from background electronic pick-up. It is apparent from the figure that as the dc breakdown current is exceeded, the power spectrum erupts from quiescence into a rich pattern of sharp spectral lines and their harmonics along with weaker broadband features. The narrow-band features can be very sharply defined: $Q$-factors of 25 are typical but values in excess of 1000 have been seen. 

It is evident from Fig. 2b that the noise ``fingerprint'' is strongly dependent upon the dc current flowing through the sample. Indeed, in some respects the noise power spectrum mirrors the dc \IV\ characteristic: Immediately above threshold \Vdc\ rises linearly with \Idc; at the same time, the frequencies of the dominant noise modes also rise roughly linearly.  At higher dc currents, \Vdc\ exhibits sharp step-like features and ultimately begins to fall. Likewise, the power spectrum undergoes abrupt structural changes and eventually the dominant mode frequencies also begin to subside. 

The slight differences between the dc \IV\ data in Figs. 2a and 1b, which were taken at the same temperature and magnetic field, are typical of the day-to-day variations encountered during the course of the experiment. Similarly, the noise spectrum exhibited variations from measurement to measurement under identical conditions as well as between measurements taken at slightly different filling factors within a given RIQHE or with different voltage contact configurations. In spite of these quantitative differences, the qualitative features of the noise, i.e. its onset near dc breakdown, its strong current dependence, and its complex spectral makeup, are robust. 

Noise features similar to those seen in the RIQHE near $\nu \approx 4+1/4$ have
also been found in several other RIQHE states. In addition to the RIQHE
near $\nu \approx 4+3/4$ in the same spin branch of the $N$=2 LL, narrow
band noise has been observed in RIQHE states near $\nu \approx 5+1/4$
and 6+1/4, the last being in the $N$=3 LL. Although the
strength and detailed spectral make-up of the noise vary widely, the
qualitative effect is the same. Figures 2c and 2d show representative
\IV\ data and noise spectra from the RIQHE near $\nu \approx 6+1/4$.

\begin{figure}
\includegraphics[width=3.25 in, bb=104 249 486 586]{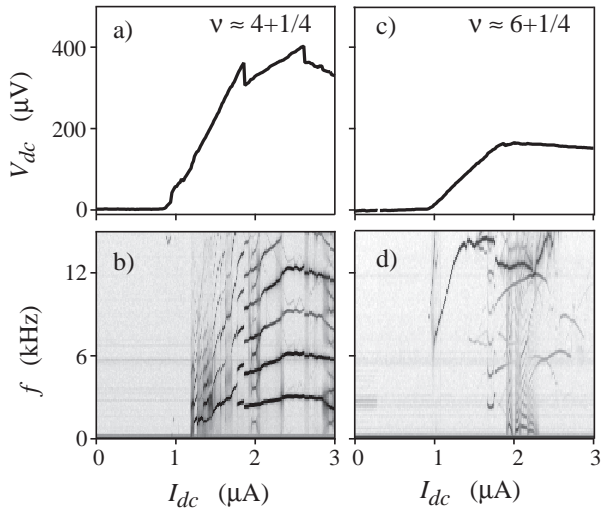} 
\caption{\label{fig:fig2} Spectral characteristics of noise associated with RIQHE breakdown.
a) \& b) \IV\ characteristic and noise power spectra at $\nu=4.29$ and $T=50$ mK.
c) \& d) \IV\ characteristic and noise power spectra at $\nu=6.24$ and $T=50$ mK.
}
\end{figure}

The noise accompanying RIQHE breakdown is restricted to the same narrow
ranges of filling factor and low temperatures as the sharp features in
the dc \IV\ curves\cite{cooper1}. In Fig. 3 the average noise level,
\Vavg, in the filling factor range $5>\nu>4$ is overlaid on the
longitudinal resistance data shown in Fig. 1a. The quantity \Vavg\
represents the rms noise voltage in a bandwidth of 1 to 25 kHz,
calculated from the noise spectra at fixed dc current and averaged over
all \Idc\ from 0 to $3~\mu {\rm A}$. Qualitatively, \Vavg\ is just the
averaged intensity of a noise fingerprint like the one in Fig.~2b. The
baseline of \Vavg\ in Fig.~3, about $1.4 ~\mu {\rm V}$, is equal to the
background noise signal in the 1-25 kHz bandwidth at \Idc$=0$. Clearly,
excess noise above this baseline is absent over most of the range
$4<\nu<5$. In particular, no excess noise is observed in the integer QHE
regions at $\nu=4$ and $\nu=5$, nor in the region of anisotropic
transport around $\nu=9/2$. (The latter is true regardless of the
direction of the current flow.) In contrast, a strong peak in \Vavg\ is
observed in the RIQHE region around $\nu \approx 4+1/4$, and a lesser but still
noticeable peak is present in the RIQHE near $\nu \approx 4+3/4$. The
substantial difference in the noise power exhibited by these two RIQHE
states is not understood. We emphasize, however, that such apparent
breakdowns of particle-hole symmetry in 2D electron systems are
commonplace. For example, recent studies of the subtle FQHE\cite{pangap}
and insulating phases\cite{eisenstein} in the $N$=1 LL have revealed a
similar lack of symmetry about half-filling. As in the present
situation, the collective states on the high magnetic field side of
half-filling are stronger than their counterparts on the low field side.
Spin effects and/or LL mixing may be involved, but there is as
yet no real understanding of such phenomena.

The inset to Fig. 3 shows that the excess noise observed in the RIQHE is
strongly temperature dependent, disappearing above about 100 mK just
like all other signatures of the recently discovered collective
phenomena in high LLs\cite{jpesscom}. 
 
\begin{figure}
\includegraphics[width=3.25 in,bb=0 0 391 239]{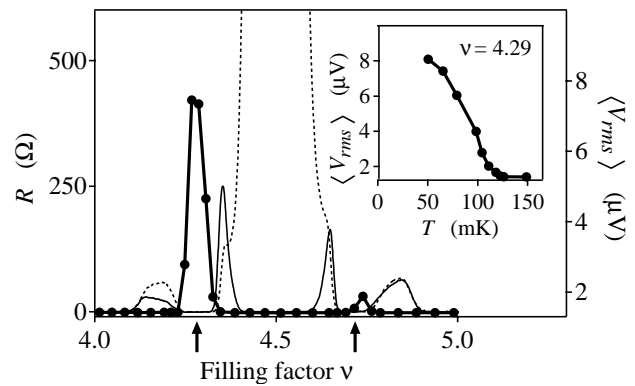}
\caption{\label{fig:fig3} 
Comparison of linear response longitudinal resistances $R$ and average noise voltage \Vavg\ for $4<\nu<5$ at $T=50$ mK. Light solid trace: $R$ with current directed along \xdir. Dashed trace: $R$ with current directed along \ydir. Dark trace with dots: \Vavg.  Arrows indicate RIQHE states. Inset: Temperature dependence of \Vavg\ at $\nu=4.29$.
}
\end{figure}

Because the breakdown of the RIQHE may reflect the depinning of an electron bubble lattice, it is logical to compare the observed narrow-band noise with the ``washboard'' noise encountered in conventional CDW compounds\cite{fleming}. Washboard noise is believed to reflect the spatial periodicity of the depinned CDW as it slides across the impurity potential in the sample.  Consistent with this, the noise frequency is found to be proportional to the current carried by the CDW\cite{gruner}. In the present case of a 2DES, the generally increasing frequency of the narrow-band noise immediately above the RIQHE breakdown is suggestive of a similar mechanism.  In addition to washboard noise, conventional CDW systems exhibit a variety of interference phenomena between narrow-band noise modes and externally applied ac signals\cite{gruner}. Indeed, we have recently observed similar effects in RIQHE breakdown, including examples of mode-locking and mode-mixing\cite{future}.

Despite these similarities, it is difficult to understand the very low
frequencies of the noise observed in our experiment. Assuming, for
example, that 100 nA flows via the sliding of a bubble CDW with a 100 nm
lattice constant, the expected washboard frequency in our sample
(assuming the current density is uniform) would be on the order of 10
MHz, not the few kHz that we observe. There are, of course,
uncertainties in such an estimate. For example, the fraction of the
total current \Idc\ which flows in the partially filled $N=2$ Landau
level is not accurately known. Below breakdown \Idc\ flows entirely
within the $N=0$ and $N=1$ LLs beneath the Fermi level. Far above
breakdown, or, equivalently, at very high temperatures, at most $\sim
6\%$ (=0.25/4.25) of \Idc\ at $\nu = 4+1/4$ flows in the valence LL.
Immediately above breakdown the situation is less clear. Hall voltage
measurements suggest that the fraction of \Idc\ flowing in the valence
LL increases smoothly from zero at breakdown to a few percent
well above it, but it is difficult to analyze such data in a
model-independent way. Although this current shunting effect could
reduce the expected noise frequencies substantially, it seems unlikely
that it can account for the kHz noise frequencies which are still
observed when the \Idc\ exceeds breakdown by hundreds of nA.

The observation of anomalously low-frequency narrow-band noise does not
rule out a sliding charge density wave picture of RIQHE breakdown. In
fact, the low bandwidth ($\sim100 ~\rm{kHz}$) of our current
experimental setup leaves higher frequency ``conventional'' narrow-band
noise undetectable in any case. Nonetheless, the existence of
low-frequency noise demands a more complex model than an ideal bubble
lattice uniformly sliding across the macroscopic dimensions of our
samples. Such a more realistic scenario will have to incorporate the
well-known inhomogeneity of current flow in the quantum Hall regime and
the possibility of non-uniform plastic or filamentary flow within the
driven CDW lattice. The latter phenomena are thought to occur in other
2D systems, such as driven vortex lattices\cite{vortex}, and have been
discussed in the context of the depinning of the hypothesized Wigner
crystal in the lowest LL at very low filling
factors\cite{fertig,olson}.

The noise reported here might instead be related to the processes responsible for the breakdown of the conventional integer QHE\cite{nachtwei}. The phenomenology of QHE breakdown is quite rich, with substantial differences between samples of different density, mobility and geometry.  Numerous mechanisms\cite{nachtwei} for the breakdown have been proposed, including runaway heating, Zener tunneling, inter-LL transitions, backscattering filament formation, and magneto-exciton generation\cite{eaves}. Not surprisingly, no single theory satisfactorily explains all the experimental results. While the breakdown of the RIQHE resembles that of the conventional integer QHE, there are important differences.  For example, the observed threshold fields for conventional integer QHE breakdown in comparable bulk 2D samples exceed those we find in the RIQHE by large factors, typically $10^3-10^4$.  In addition, narrow-band noise has not been observed in conventional integer QHE breakdown. 

In conclusion, we have observed complex noise spectra accompanying the
breakdown of the re-entrant integer quantized Hall states of 2DESs 
in high LLs. These states, which represent insulating
configurations of the electrons in the uppermost LL, may well
be pinned ``bubble'' charge density waves. After breakdown under the
stress of a large Hall electric field, both narrow- and broad-band noise
appears in the current-voltage characteristics of these states. Although
the frequencies of the narrow-band modes generally increase with the
current driven through the system, they are too low to be readily
accounted for by a simple sliding charge density wave model.

We thank V. Oganesyan, S. Brown, and R. Thorne for useful discussions and M.P. Lilly for his work on the early stages of this experiment.  This work was supported by the DOE under Grant No. DE-FG03-99ER45766 and the NSF under Grant No. DMR0070890.

\end{document}